# ON THE THERMAL CONSOLIDATION OF BOOM CLAY

P. Delage, N. Sultan and Y.J. Cui

Ecole Nationale des Ponts et Chaussées
Navier/CERMES (Geotechnical department)
6-8 Av. B. Pascal,
F - 77455 Marne le Vallée cedex 2
France

**ABSTRACT** : When a mass of saturated clay is heated, as in the case of host soils surrounding nuclear waste disposals at great depth, the thermal expansion of the constituents generates excess pore pressures. The mass of clay is submitted to gradients of pore pressure and temperature, to hydraulic and thermal flows, and to changes in its mechanical properties. In this work, some of these aspects were experimentally studied in the case of Boom clay, so as to help predicting the response of the soil, in relation with investigations made in the Belgian underground laboratory at Mol. Results of slow heating tests with careful volume change measurements showed that a reasonable prediction of the thermal expansion of the clay-water system was obtained by using the thermal properties of free water. In spite of the density of Boom clay, no significant effect of water adsorption was observed. The thermal consolidation of Boom clay was studied through fast heating tests. A simple analysis shows that the hydraulic and thermal transfers are uncoupled. Experimental results from fast heating tests showed that the consolidation coefficient does not change significantly with increased temperature, due to the opposite effect of increasing permeability and decreasing porosity. The changes of permeability with temperature were investigated by running constant head measurements at various temperatures. An indirect analysis, based on the estimation of the $m_v$ consolidation parameter, showed that the indirect method of estimating the permeability from consolidation tests should be considered carefully. Intrinsic permeability values were derived by considering the change of the viscosity of free water with temperature. A unique relationship between the intrinsic permeability and the porosity was observed, with no dependence on temperature, confirming that the flow involved in the permeability test only concerns free water.

**RESUME** : Quand un massif d'argile saturée est soumise à une élévation de température, comme dans le cas des sols hôtes pour les stockages nucléaires à grande profondeur, la dilatation thermique des constituants engendre des surpressions interstitielles. Leur dissipation implique des gradients de pression et de température, des flux d'eau et de chaleur, et une modification des propriétés mécaniques du sol. Certains de ces aspects sont étudiés ici dans le cas particulier de l'argile de Boom, afin de contribuer à l'étude de la réponse du sol, en relation avec les recherches effectuées au laboratoire souterrain de Mol en Belgique. Les résultats d'essais de chauffage lent avec une mesure soigneuse des variations de volume ont montré qu'une prévision raisonnable de la dilatation thermique de l'eau interstitielle était obtenue en utilisant les coefficients thermiques de l'eau libre, avec donc un effet négligeable de l'eau adsorbée contenue dans cette argile plutôt dense. La consolidation thermique de l'argile de Boom a été étudiée à l'aide d'essais de chauffage rapide. Comme le montre l'analyse, on observe que les transferts hydriques et thermiques sont découplés. Les valeurs obtenues du coefficient de consolidation sont peu variables avec la température, du fait des effets opposés de la diminution de volume et de l'augmentation de la perméabilité avec la température. Des essais de perméabilité à charge constante à différentes températures ont permis d'étudier les variations de la perméabilité avec la température, et de montrer les insuffisances de la détermination indirecte à l'aide des courbes de consolidation. En utilisant les variations de la viscosité avec la température de l'eau libre, on observe une relation unique entre la perméabilité intrinsèque et la porosité, qui confirme que l'eau mobilisée pendant un essai de perméabilité est l'eau libre.

# INTRODUCTION

The effect of temperature on the behaviour of soils is a crucial problem in geoenviromental engineering. Soils surrounding nuclear waste disposal or buried high voltage cables are exposed to an elevated temperature during a long period of time, and suffer changes of their hydraulic and mechanical properties. In saturated soils, heating induces thermal expansion of the pore water and of the soil solids, resulting in the build-up of thermal pore pressures, reduction of the in-situ effective stress, and subsequent volume changes due to the dissipation of pore pressure and heat transfer. The dissipation of thermal induced pore pressure with time is a complex phenomenon, involving gradients of pore pressure and temperature, hydraulic and thermal flows within the mass of soil, and changes in the mechanical properties of the soil with temperature.

Particular attention has recently been paid to the thermomechanical behaviour of Boom clay, in relation with the underground laboratory of the Belgian Nuclear Research Center (SCK-CEN) at Mol, which is located in the Boom clay deposit. The investigations to date mainly have dealt with the effect of temperature on the mechanical properties of the clay, and they were primarily based on the results of drained tests. The dissipation of temperature-induced pore pressures has not been specifically addressed in the case of Boom clay.

A review of the literature shows that the various physical phenomena involved in thermal consolidation are poorly documented. Few experimental data deal with the thermal expansion of the clay-water system, which is the phenomenon generating excess pore pressures in a mass of clay heated in-situ. More available data concern standard consolidation tests at different constant temperatures, and drained heating tests under a constant total stress. Various authors derived permeability values from consolidation curves, whereas very few direct permeability measurements at different temperatures have been carried out. The objective of this paper is to present an experimental study dedicated to Boom clay, dealing with the thermal volume change of the clay-water system, the thermal consolidation, and the effect of temperature on permeability. The study is aimed at better predict the response of a host soil to the heat generated by nuclear wastes.

# BACKGROUND

The early studies on the effect of temperature on soil behaviour were related to the temperature change occurring within a soil sample, when sampled and transported to the laboratory. Gray (1936) performed isothermal oedometer tests at 10°C and 20°C, and showed a difference in terms of compressibility and preconsolidation pressure. Paaswell (1967) carried out heating tests under a constant load in a rigid ring oedometer. He observed that the curves of volume changes as a function of time were similar to that of standard consolidation, leading to the concept of thermal consolidation. In thermal consolidation, excess pore pressures are generated by the differential thermal expansion of the soil constituents (solid+water). Campanella and Mitchell (1968) performed similar tests in an isotropic compression cell, and presented a detailed study of the thermal dilation of the soil constituents. They also investigated thermal induced pore pressure in undrained samples. Habibagahi (1977) performed isothermal consolidation tests at various temperatures in a rigid ring oedometer, derived permeability values at different temperatures, and commented about temperature effect on the consolidation coefficient and on the permeability. He showed that permeability increased with increased temperature, due to the decrease in water viscosity. This was confirmed by similar tests carried out by Towhata et al. (1993). Houston and Lin (1987) studied the thermal consolidation of deep submarine sediments.

Direct measurements of the permeability at various temperatures are rare. Morin and Silva (1984) performed constant head permeability tests in a rigid ring cell, on various ocean sediments at temperatures ranging from 22°C to 220°C, confirming the previous conclusions. The validity of the indirect method of measuring permeability, as compared to direct methods, has been discussed in Burghignoli et al. (1995) and Towhata et al. (1995), with no clear conclusion, leaving the question opened.

Thermally induced volume changes need to be accounted for when considering permeability changes with temperature. This aspect has been well documented on various clays (Campanella and Mitchell 1968, Plum and Esrig 1969, Demars and Charles 1982, Towhata et al. 1993), including Boom clay (Baldi et al. 1988, 1991, Sultan 1997). By running isothermal compression tests at various temperatures, most authors have shown little influence of temperature on the coefficients of elastic and plastic compressibility. The elastic domain of a soil is known to reduce with increased temperatures (Tidfors and Sällfors 1989, Eriksson 1989, Boudali et al. 1994), and the drained volume change behaviour of a soil submitted to heating is depending of the overconsolidation ratio. When temperature is increased, a highly over-consolidated soil initially exhibits an expansion, followed by a contraction, whereas a normally consolidated (or lightly overconsolidated) soils only exhibit an irreversible contraction. The initial expansion of highly overconsolidated soils arise from the thermal expansion of the soil minerals, and they are reversible with temperature.

The thermal contraction is related by various authors to physico-chemical clay-water interactions, based on the change in thickness of the double layer with increased temperature. Plum and Esrig (1969) comment on the contradictory conclusions drawn from Lambe's model (1960), which predicts a depression of the double layer with an increased temperature, and from the work of Yong et al. (1962), who describes an increase of the double layer thickness with temperature, resulting in an increase of the repulsive forces. Morin and Silva (1984) consider the theoretical expression of the double layer thickness (see for instance Mitchell 1993), which shows that the thickness is proportional to the square root of temperature. They conclude that the double layer thickness should increase with increased temperature. Mitchell (1993) considers that this effect is probably counterbalanced by a decrease of the dielectric constant, and that the double layer thickness should not change significantly with temperature.

Towhata et al. (1993) noted that the double layer theory is essentially based on reversible phenomena, and could not account for the irreversible thermal contraction of normally consolidated samples. Paaswell (1967) considered that heat would cause a thermal agitation of the water molecules bound to the clay particles, enabling them to move out of the bound layer with greater ease. When water molecules are expelled, the thickness of the bound layer reduces, and clay particles can get closer one to another, resulting in an irreversible volume decrease. Habibagahi (1977) considered that the molecules of adsorbed water do not move during a permeability test at a given temperature, and that the flow of free water occurs through channels limited by the clay particles and the adsorbed water. When temperature is increased, the thickness of the bound water layer decreases, resulting in larger channels, and a higher permeability. This analysis is consistent with the decrease in plastic limits with increased temperature, evidenced by Youssef et al. (1961).

In fine grained soils, the distinction between bound water and free water is somewhat schematic, due to the progressive decrease in tightness of the clay-water link with increased clay-water distance. This was interpreted by Low and Lovell (1959) (in Towhata et al. 1993) in terms of an apparent viscosity, larger close to the clay particles, and decreasing with the clay-water distance. In this regard, it is difficult to clearly separate viscosity effects from those related to clay-water interaction, when analysing the effect of temperature on water flow in clayey soils. As a consequence, the use of the standard exponential expression giving the change of viscosity of free water with temperature is approximate.

In order to predict the thermal expansion of a saturated soil submitted to an increased temperature, Campanella and Mitchell (1968) considered the thermal properties of free water. The thermal expansion $\Delta V_w$ of a volume of pore water submitted to a temperature increment $\Delta T$ is equal to :

$$\Delta V_w = \alpha_w V_w \Delta T \tag{1}$$

where $\alpha_w$ is the thermal expansion coefficient of free water. The corresponding volumetric strain $\varepsilon_v$ during a drained heating test is obtained by subtracting the volume due to the thermal expansion of water and solids from the total volume of drained pore water $\Delta V_{dr}$, giving :

$$\varepsilon_v = \frac{\left[\Delta V_{dr} - (\alpha_w V_w + \alpha_s V_s)\Delta T\right]}{V} \quad (2)$$

where $\alpha_s$ and $V_s$ are respectively the thermal expansion coefficient and the volume of the solid skeleton, and $V$ the volume of the sample.

In low porosity plastic clays, Baldi et al. (1988) accounted for the effects of adsorbed water using the double layer theory. They considered that the molecules of adsorbed water are submitted to a pressure exponentially decreasing with the distance between the water molecule and the clay platelet, and accounted for the effects of temperature and pressure on the volume change of the adsorbed water. Considering that there is no free water in dense plastic clays, Baldi et al. (1988) finally obtained the following expression of the thermal volumetric strain :

$$\varepsilon_v = \frac{\left[\Delta V_{dr} - (V\Delta v_a S_s \rho_d + \alpha_s V_s)\Delta T\right]}{V} \quad (3)$$

where $\Delta v_a$ is the volume of expanded adsorbed water per unit surface of clay mineral and per °C, $S_s$ the specific surface of the clay and $\rho_d$ the dry unit weight of the soil. For Boom clay, Baldi et al. (1988) provided the following parameters :
$\delta v_a = (401.99 + 1.2897\,T)\,10^{-14}$ (m/°C), $\quad S_s = 180$ m$^2$/g, $\rho_d = 19$ Mg/m$^3$, $w = 24\%$, $n = 37.5\%$, $\alpha_s = 10^{-5}$ (°C). In this study, these values are adopted.

**EXPERIMENTAL SET-UP**

Tests were performed on samples, extracted from a depth of 223 m in the Boom clay deposit (Decleer et al. 1983), in the underground facilities of SCK-CEN in Mol (Belgium). The thermo-mechanical behaviour of Boom clay has been experimentally studied by various authors (Baldi et al. 1988, 1991, Neerdael et al. 1992, De Bruyn and Thimus 1995, Bernier et al. 1995, Belanteur et al. 1997), in relation with nuclear waste disposal. Boom clay is a stiff clay, with a plasticity index of about 50%, a natural porosity around 40% and a water content which varies between 24 and 30%.

Tests were carried out in an isotropic compression cell (see Figure 1) designed to support high pressures and high temperatures (up to 60 MPa and 100°C respectively). A heating coil was placed on the outer wall of the cell, and the temperature was controlled by a thermocouple placed inside the cell, in the confining water. The precision of the temperature regulation system is ±0.05°C. The isotropic stress was applied by a high pressure controller (GDS type, 60 MPa), and the back pressure (up to 2 MPa) by a standard GDS controller. The main advantage of this system is its ability to monitor volume changes while applying pressures.

In standard isothermal triaxial tests on saturated samples of soils, volume changes are measured by monitoring the transfer of pore water to or from a test specimen. This is no longer valid in non-isothermal conditions, because of the thermal dilation of water and solids. For this reason, Baldi et al. (1988) calculated volume changes from the measurements of axial strains, under an hypothesis of isotropic soil. Here, the feasibility of volume change measurements based on the monitoring of the transfer of water to or from the confining cell has been considered.

In standard triaxial tests (confining pressure smaller than 1.5 MPa), volume change measurements by monitoring the exchanges of confining water are not valid, due to excessive perturbations created by the compression of various devices (plastic tubing, perspex cell wall, membranes - see for instance Cui, 1993). In our case, most components of the high pressure cell are metallic and much stiffer, with a repeatable and reversible response to pressure cycles. For this reason, the calibration of the system allows a satisfactory volume change measurement, with a good agreement between inner and outer measurements (Figure 2). In addition, it should be noted that heating tests under a constant cell pressure are not affected by this phenomenon, since no compression of the internal devices occurs.

A careful calibration of the volume change measurements during compression was made on a saturated sample of Boom clay at 20°C. The quantity of water introduced during the compression test by the cell pressure controller (outer measurement) was compared to the amount of pore-water expelled from the sample, and monitored by the back pressure controller (inner measurement). Figure 2 shows that the two curves are reasonably similar. The correspondence is quite good in the first loading stage, up to 1 MPa. A larger difference is observed during the second loading stage, providing an uncertainty lower than 0.01 point on the void ratio between 1 and 4 MPa.

The accuracy of volume change measurements during heating tests are affected by the thermal expansion of the device. This also concerns the changes in the temperature gradients in the water located in the internal ducts connecting the controllers (at the ambient temperature, i.e. 20°C) either to the base of the sample, or to the water cell (both at test temperature). In order to minimise this problem, the metal tubes connecting the cell to the pressure controllers were immersed in water baths maintained at 20°C (see Fig. 1). For low heating rates and slow movements of water during heating, the system imposes the 20°C temperature condition right at the outlet. Thereby the volumes of the water located in the cell base, in which temperature is decreasing, are maintained constant. The effect of the thermal dilation of these volumes of water during heating was calibrated at a same rate of heating, under various cell pressures, using a metal sample of a known thermal dilation coefficient.

The thermal calibration of the device was carried out by performing tests along various paths of pressure and temperature. Heating tests (between 20 and 100°C) were carried out on a blank metallic sample under 6 distinct values of the all round pressure (1, 2, 3, 5, 10 and 12 MPa). Compression tests between 0 and 10 MPa were carried out under temperatures of 20, 80 and 100°C.

**THERMAL VOLUME CHANGE OF THE CLAY WATER SYSTEM**

The two theories on the thermal dilation of the clay-water system, presented in the Background section, are now examined by running slow heating tests. Four samples of natural Boom clay were saturated under a back pressure of 1 MPa and an isotropic effective stress of 30 kPa. After one week, a value of 0.97 of the B Skempton coefficient was obtained, showing a satisfactory saturation. The samples were slowly consolidated up to an all round pressure of 4 MPa, using a stress rate of 40 kPa/h. This rate was found acceptable by checking constantly the value of the pore pressure, which remained equal to 1 MPa. The samples were then slowly heated (0.3°C/h) up to 100°C. This rate was also found to be slow enough by checking that the pore pressure value remained constant, ensuring a constant isotropic effective stress of 4 MPa during heating.

Four different heating tests were performed in order to check the repeatability of the results obtained on the different intact natural samples. Figure 3 shows the results in terms of porosity and volume of drained water versus the temperature. Some difference in the values of the initial porosity are observed. The curves of drained water are starting horizontally, showing little expulsion of pore water below 40°C. Above 40°C, the slope of the curve is increasing significantly, showing more effect of heating at higher temperatures.

The porosity curves of samples 2, 3 and 4 all present an initial decrease between 20 and 30°C. The effect is slight in curve (4), it increases in tests (3) and (2). This effect is not consistent with the horizontal slopes of the curves of drained water, which are representative of the volume change behaviour of the sample at lower temperatures. In this regard, the quality of test (1) at the beginning of heating seems to be more reasonable, since the initial slope is horizontal. Hence, the initial loss of porosity of tests 2 to 4 is believed to be artificial, and related to the initial adjustments of the system (which however could not be completely identified). One considers that little volume change occurs below 40°C. In the following, the initial loss of porosity is not considered anymore, and corrected volume change curves are obtained by extrapolating the horizontal plateau observed between 30 and 55°C at the beginning of the curves.

The experimental curves of volume changes are now compared (Figure 4) with the volume changes curves derived from water exchanges according to equation (2) (Campanella and Mitchell 1968) and (9) (Baldi et al. 1988). The curves of drained water are also shown. It is observed that the quality of the predictions apparently depend on the test. In test 1, poor predictions are observed below 70°C, and the adsorbed water prediction fits well above this temperature. In test 2, little difference is observed between the two predictions, and both overestimate the contraction at temperatures higher than 70°C. In test 3, a significant difference is observed between the two predictions, and the experimental curve is comprised between the two predictive curves, close to Baldi et al.'s prediction. In test 4, both predictions only slightly underestimate the experimental response.

In Figure 5, the measured thermal dilation of the soil constituents is obtained by subtracting the sample volume change ($\varepsilon_v V$) from the volume of drained water ($\Delta V_{dr}$), as explained by Campanella and Mitchell (1968). Experimental results are compared to results given by the theories of Campanella and Mitchell, and Baldi et al., corresponding respectively to :

$$\Delta V_{dr} - \varepsilon_v V = (\alpha_w V_w + \alpha_s V_s)\Delta T \qquad (4)$$

and

$$\Delta V_{dr} - \varepsilon_v V = (V \Delta v_a S_s \rho_d + \alpha_s V_s)\Delta T \qquad (5)$$

Equation (10) only depends on the volumes $V_w$ and $V_s$, showing that the free water prediction is only depending on the void ratio. The same observation is true for the adsorbed water prediction (Eq. 11), as $\Delta v_a$ is similar in all the samples. The sensitivity of the predictions regarding the value of the initial void ratio for the four samples is examined in Figure 5.

In Campanella and Mitchell's expression, the $\alpha_w$ coefficient is constant, and no influence of the water pressure on water thermal dilation is accounted for. Since the tests are carried out with a back pressure of 1 MPa, it was decided to adopt an expression given by Baldi et al. 1988, which gives $\alpha_w$ as a function of temperature and pressure as follows :

$$\alpha_w(T,p) = \alpha_0 + (\alpha_1 + \beta_1 T) \ln mp + (\alpha_2 + \beta_2 T)(\ln mp)^2$$

This value of $\alpha_w(T,p)$ has been used in equation (10), and the corresponding points are also presented in Figure 5 (Modified free water).

Various conclusions may be drawn from the figure : a) the effect of the natural variability of the void ratio is not very significant; b) both predictions overestimate the expansion below 60°C, but the difference between the two predictions is not significant as compared to experimental dispersion; c) experimental points are located between the free water and adsorbed water predictive curves at temperatures larger than 60°C; d) the predictions given by the modified Campanella and Mitchell's approach, with a coefficient of thermal expansion of water changing with the pore pressure, are included between those of the two other models.

If one considers the natural variability of clays, the experimental dispersion, and the difficulty of determining some microstructure parameters which characterise the status of bound water in the model of Baldi et al., it can be concluded that the simpler model of Campanella and Mitchell is satisfactory. It main advantage is that it is based on the thermal and compression properties of free water, with no influence of the clay parameters.

**THERMAL CONSOLIDATION TESTS**

A sample of natural Boom clay was saturated under a back pressure of 1 MPa and an isotropic effective stress of 30 kPa. Once saturated, the soil was isotropically consolidated in the temperature controlled laboratory (20°C) under an all round pressure of 4 MPa, and subsequently unloaded down to 2 MPa, leading to an overconsolidation ratio $OCR = 2$.

Thermal consolidation tests at different temperatures were carried out on the sample by applying 10°C temperature increments. Starting from 20°C, the temperature regulating system was switched on at the desired temperature (30°C in the first heating step), till stabilisation at the desired temperature was detected by the thermocouple placed in the water inside the confining cell (see Figure 1). The duration of the heating phase depends on the power of the heating system, on the mass and on the heat capacity of the system. It lasted approximately 1.5 hours. During the test, the 1 MPa back pressure was maintained using the GDS controller connected to the base of the sample, whereas the top connection was closed. So, the length of drainage was equal to the length of the sample. The test can be interpreted like a standard oedometer test, and two kinds of curves are obtained : a) curves giving the volume changes as a function of time, which are related to the rate of dissipation of excess pore pressures, and b) the strain-temperature curves, obtained by considering all final points, after stabilisation of the volume change has been attained.

Figure 6 shows the thermal volume changes curves versus the logarithm of time of the tests at low temperatures (from 23°C up to 50°C). For these tests, a final expansion is observed. Between 23 and 30°C, a peak corresponding to initial thermal expansion is observed after 2 hours, followed by a dissipation phase and a contraction smaller than the initial expansion. In spite of some oscillations which make things less clear, the same phenomenon is observed on the two other curves, resulting also in a final expansion. Due to the small amplitude of the curves and to the oscillations, it has been preferred to study the thermal consolidation by considering the contracting volume changes observed at higher temperatures.

Figure 7 shows the 4 consolidation curves obtained at higher temperatures (50-60, 60-70, 70-80 and 80-95°C), where a larger contraction is following an initially smaller thermal expansion. The expansion corresponds to the thermal expansion of the solid and water phases, which is lasting approximately 2.2 hours, due to the rate of heating adopted. In the contraction stage, the shapes of the curves are similar to that of standard consolidation curves, and show that the volume decrease is related to pore pressure dissipation.

Observation of the curves of the tests 50 - 60°C and 80 - 95°C, which have been longer (55 and 40 hours respectively) show that the secondary consolidation can reasonably be neglected, even at higher temperature. This is related to the low porosity of the clay.

Figure 8 presents the same results, in a diagram giving the relative volume change versus the logarithm of time. It shows that the relative magnitude of the initial expansion is continuously decreasing with increased temperature. This is due to the increase of the thermal contraction with temperature observed in Figure 7 after the first 2 hours, while the thermal expansion of the constituents, which only depends on the mass of solid (constant) and water (slightly decreasing) remains approximately constant.

In Figure 9, all final points, obtained after each stage of pore pressure dissipation of the test of Figure 7, are gathered together, giving the thermal volumetric response of the sample tested, as a function of temperature. Results are typical of over-consolidated soils, as shown by Hueckel and Baldi (1990). The sample is expanding at temperatures smaller than 50°C, and contracting at higher temperatures.

**PERMEABILITY MEASUREMENTS AT VARIOUS TEMPERATURES**

Constant head permeability measurements were carried out on a Boom clay sample at various temperatures, using the device presented in Figure 1. In order to ensure significant flow rates, a sample smaller than the standard triaxial sample was used, with a standard diameter ($\phi$ = 37.7 mm), and a height reduced down to $h$ = 22.95 mm. The sample was saturated under a low all round pressure (110 kPa), with a back pressure equal to 40 kPa. A $B$ coefficient of 0.97 was attained after 24 hours. Afterwards, the sample was consolidated up to 2.5 MPa. The first permeability measurement was carried out at T = 20°C by increasing the back pressure up to 1 MPa at the bottom of the sample, while putting the top porous stone in contact with atmospheric pressure. It results in a high value of the hydraulic gradient (i ≈ 4000), as compared to that of variable head permeameters (i ≈ 100). The high gradient was found necessary in order to create a measurable flow in the dense plastic clay, and to get a satisfactory precision in the measurement of flow rate and, hence, of the permeability.

In Figure 10, the volume of water injected at the base is given as a function of time. In spite of the high gradient applied, the figure shows that 10 hours are necessary to achieve permanent flow, and 15 hours to obtain a satisfactory determination of the slope of the curve corresponding to a constant flow, from which the permeability was calculated. At 20°C, a permeability value of 2.5 $10^{-12}$ m/s was obtained with a porosity equal to 39%. Possible problems related to the high value of the hydraulic gradient were examined by comparing this result with that of a variable head test, run in a rigid oedometer cell (i ≈ 100). A permeability value of 3.5 $10^{-12}$ m/s for a porosity of 44.1% was obtained, which is consistent with the former result (see Figure 12).

The test carried out for the determination of the permeability at various temperatures is presented in Figure 11, in a diagram giving the porosity versus the logarithm of the permeability. Constant head permeability tests were successively performed under an all-round pressure of 2.5 MPa at 20°C, 60°C, 70°C, 80°C and 90°C. Heating phases were very progressive, at a heating rate of 0.1°C/20 minutes, in order to fulfil drained conditions (Sultan 1997, Cui et al. 1999). At 90°C, the sample was isothermally loaded from 2.5 up to 4 MPa, and permeability tests were carried out in a cooling phase at 90°C, 80°C, 70°C, 60°C and 30°C. At 30°C, subsequent isothermal loading from 4 MPa up to 6 MPa was carried out, and permeability tests were performed at 30°C and 60°C.

One observes in Figure 11 that heating from 20°C up to 90°C results in a contraction of the sample (the porosity *n* decreasing from 39 % down to 37.2 %), and an increase in permeability, from 2.5 $10^{-12}$ m/s to 6.2 $10^{-12}$ m/s. The loading at 4 MPa and 90°C decreases the permeability down to 4.4 $10^{-12}$ m/s. During the cooling phase under 4 MPa, the permeability decreases from 4.4 $10^{-12}$ m/s down to 2.4 $10^{-12}$ m/s, the porosity remaining almost constant (34 % to 33.9 %). Loading up to 6 MPa under 30°C decreases the porosity down to 31.9%, and heating at 60°C under 6 MPa gives a final porosity of 30.6%.

**INTERPRETATION AND DISCUSSION**

Experimental evidence gained from slow heating tests (Figure 5), in which the amount of expelled water was compared to the volume change of the heated sample, showed that the simpler theory of Campanella and Mitchell (1968) appears to be satisfactory, provided the effect of the water pressure on the thermal expansion coefficient of water is accounted for. This conclusion concerns a dense natural clay, with a plasticity index of 50. It should not be generalised to dense clays with higher plasticity indexes, such as dense swelling clays considered for making engineered barriers for nuclear waste.

The thermal consolidation tests presented in Figure 7 are analysed by considering the two relevant equations :
- Fourier's equation of heat transfers :

$$D_T \nabla^2 T = \frac{\partial T}{\partial t} \qquad (6)$$

- Terzaghi's consolidation equation for pore pressure dissipation :

$$\frac{\partial U(z,t)}{\partial t} = c_v \frac{\partial^2 U(z,t)}{\partial z^2} \qquad (7)$$

where $c_v$ is the coefficient of consolidation, and $D_T$ the thermal diffusivity (equal to the ratio *K/C*, *K* and *C* being respectively the thermal conductivity and the volumetric heat capacity). The standard triaxial samples (radius $r$ = 19 mm and height $h$ = 76 mm) tests were drained by the bottom only (length of drainage 76 mm) and heated by the cell water all around, and by the top and bottom bases. A conservative assumption of radial heating (characteristic length equal to the radius 19 mm) was made. Calculations based on the values of $D_T$ and $c_v$ (taken from Picard 1993 : $c_v$ = 7.5 m$^2$/s, $K$ = 1.7 W/K/m and $C$ = 2.85 10$^-$ J/K/m$^3$, giving $D_T$ = 5.96 10$^{-7}$ m$^2$/s) and on the characteristic lengths of heat transfer and pore water dissipation, showed that heat equilibrium was reached much faster (about 10 minutes) than pore pressure dissipation (about 21.5 hours). It confirms that the two phenomena can reasonably be uncoupled, as indicated in Figure 7. In other words, there is a negligible amount of water drained during the heating phase, which lasts for 1.5 hours. In this phase, the volume increase is due to the thermal undrained expansion of water and minerals. Afterwards, the temperature of the sample is uniform, and the tests is similar to a consolidation test at a constant temperature.

Very few experimental results concerning the combined effect of thermal dilation and pore pressure dissipation during thermal consolidation similar to that of Figure 7 are available in the literature. Conversely, the problem has been modelled by using FEM numerical resolutions, within a thermo-poro-elastic framework (Aboustit et al. 1982; Noorishad et al. 1984; Lewis et al. 1986; Gatmiri and Delage 1997). All authors considered the thermal consolidation of a sand, in which the situation is opposite to that of Figure 7 : since $c_v$ is much smaller than $D_T$, water is drained first, resulting in a contraction, and the thermal expansion of the grains occurs afterwards (see Figure 12, taken from the data of Noorishad et al. 1984). In the lack of relevant experimental data, all FEM calculations were validated by comparing with an analytical solution given by Aboustit and al.. In this regard, it is thought that the experimental results of Figure 7 can be used as a physical validation case for these codes, in the case of the thermal consolidation of a fine-grained soil.

Changes in the consolidation coefficient $c_v$ with temperature were deduced from curves of Figure 7, considering the points at $t_{50}$. Results are reported in Figure 13. The curves show no significant change in the value of $c_v$ with temperature, with an increase from 3.43 10$^{-8}$ m$^2$/s up to 4.32 10$^{-8}$ m$^2$/s when temperature is increased from 60 to 70°C, followed by a plateau between 70 and 95°C. This variation is smaller than the order of magnitude, as observed by Habibagahi 1977 on a low plasticity clay ($I_p$ = 25, at $T$ = 25 and 50°C) and Towhata et al. 1993 on two clays ($I_p$ = 27 and 42, at $T$ ranging from 25 to 200°C). The slight changes of $c_v$ observed in Figure 13 is related to two opposite simultaneous effects : the increased temperature increases the permeability of the sample, but this effect is compensated by the decrease of the void ratio.

Most existing data on the effect of temperature on the permeability of clays (Habibagahi 1977, Towhata et al. 1993) come from $c_v$ measurements obtained from isothermal consolidation tests performed at various temperatures. This technique appears somewhat questionable (Burghignoli and al 1995). Due to technical difficulties, no pore pressure measurements have been carried out during the heating tests presented here, and it is not possible to derive any permeability value from $c_v$ measurements. However, taking into account the relation between $c_v$ and $k$ :

$$k = m_v \, \gamma_w \, c_v \qquad (8)$$

it is possible to evaluate a value of $m_v$, and to compare it to the experimental compression curve of the thermal consolidation test. The values of $c_v$ obtained from the consolidation curves of Figure 7 are combined to permeability values deduced from Figure 11, in order to evaluate the values of $m_v$. Results presented in table 1 show that the calculated values of $m_v$ are from 3 to 4 times larger than those obtained from the compression test, at a same porosity. This shows that indirect permeability measurements from consolidation tests should be considered carefully, since they apparently overestimate the permeability. Obviously, this point would deserve a specific experimental study with pore pressure measurements, to be carried out on a variety of soils, and the preliminary results presented here on Boom clay should be completed by an extensive program, similar to that conducted by Morin and Silva (1984).

The changes in permeability presented in Figure 11 are due to the coupled effect of changes in temperature and porosity. In order to separate these two effects, the intrinsic permeability values ($K$) were calculated according to :

$$k = \frac{K\gamma_w(T)}{\mu(T)} \qquad (9)$$

where $\mu(T)$ is the water viscosity and $\gamma_w(T)$ is the unit weight of water. The following relation, valid for free water, was derived from experimental values reported by Hillel (1980) :

$$\mu(T) = -0.00046575\ln(T) + 0.00239138 \quad (Pa.s) \qquad (10)$$

In the range of temperatures considered, there is no significant change in the value of $\gamma_w(T)$.

Observation of the results presented in Figure 14 in a semi-logarithmic diagram shows that the relationship between the intrinsic permeability $K$ and the porosity $n$ is only and linear, and appears reasonably independent of temperature. The data obtained from the variable head test performed in the rigid ring oedometer correctly matches with the results. Consequently, the intrinsic permeability of a sample loaded at a given temperature is only dependent on its porosity, independently on the thermo-mechanical path previously followed in a ($p'$;$T$) plane. In other words, volume changes created by stress and/or temperature changes have the same effect on the intrinsic permeability of a sample. Data of Figure 14 are also consistent with the interpretation of Habibagahi, and confirm that the water put in movement during a permeability test is the free water, which circulates through the channels limited by the water tightly adsorbed on the clay minerals.

Similar conclusions regarding the intrinsic permeability were drawn by Morin and Silva (1984), from permeability tests carried out at various temperatures (between 22 and 220°C) on 4 soils of different plasticity indices and higher void ratios ($I_p$ between 52 and 179, and $e$ between 1 and 9). All the permeability values obtained by Morin and Silva at various temperatures and stresses are presented in Figure 15, together with the data of Figure 12 on Boom clay ($I_p = 50$). The figure shows that the respective position of the curves is governed by the plasticity index. In this regard, the position of the points of Boom clay is consistent with the results obtained from the two soils of a similar plasticity index ($I_P = 52$ and 59). It confirms that, at a given void ratio, a more plastic soil will be more impervious, due to a greater number of layers of water molecules adsorbed on the clay minerals, which do not participate to the flow, as commented by Habibagahi (1977).

**CONCLUSIONS**

The following concluding remarks can be derived from the study :

- a) For Boom clay, the experimental study of the thermal expansion of the clay-water system shows that a satisfactory prediction of the thermal expansion of the pore water is obtained with the simpler free water model of Campanella and Mitchell (1968), provided the influence of the pore water pressure on the thermal expansion coefficient of water is accounted for. This remark is valid for the samples of Boom clay ($I_p = 50$) studied herein. It should be confirmed for any higher plasticity or denser clay considered for nuclear waste disposal, in order to investigate any possible effect of adsorption, as suggested by Baldi et al. (1988).

- b) Slow heating tests carried out on a slightly overconsolidated sample of Boom clay showed that, in the contracting phase, i.e. for temperature higher than 60°C and up to 90°C, the change of the coefficient of consolidation $c_v$ during thermal consolidation is not significant. The effects of the decrease in porosity are compensated by the increase in permeability with temperature. This simple conclusion, which confirms previous works and extend them to the case of Boom clay, is of interest for the prediction of the dissipation of the thermal pore pressures in the vicinity of nuclear waste disposals in clays at great depth.

- c) Investigations on the effect of temperature on the permeability should be made through direct permeability tests at various temperatures, since it has been observed that results from consolidation curves overestimate the permeability by a factor of about 4, in the case of Boom clay.

-d) As observed by Morin and Silva on various clays, there is an unique and linear relationship between the porosity and the logarithm of the intrinsic permeability of Boom clay, independently of the thermo-mechanical path previously followed by the sample. Consequently, 1) the decrease in permeability of a sample with temperature is only related to the decrease in the viscosity of water and 2) the water moved during a permeability test is free water.

## ACKNOWLEDGEMENTS

The authors are indebted to MM. Neerdael and De Bruyn, from SCK-CEN at Mol (Belgium), for providing the samples of Boom clay. This work has been developed in relation with the European Human Capital and Mobility Network ALERT. It is part of the second author's PhD thesis, which was supported by a grant of Ecole Nationale des Ponts et Chaussées. The comments made by the reviewers during the evaluation of the paper were greatly appreciated.

## REFERENCES

Aboustit, B.L., S.H. Advani, J.K. Lee & R.S. Sandhu 1982. Finite element evaluations of thermoelastic consolidation. *Issues on Rock Mechanics, 23rd Symp. on Rock Mechanics, ASME, Univ. of California-Berkeley* : 587-595.

Baldi, G, T. Hueckel & R. Pellegrini 1988. Thermal volume changes of the mineral-water system in low-porosity clay soils. *Canadian Geotechnical Journal* **25** : 807-825.

Baldi, G, T. Hueckel, A. Peano & R. Pellegrini 1991. Developments in modelling of thermo-hydro-geomechanical behaviour of Boom clay and clay-based buffer materials (vol. **2**). *Commission of the European Communities, Nuclear Science and Technology* EUR 13365/2.

Belanteur, N., S. Tacherifet & M. Pakzad 1997. Etude des comportements mécanique, thermo-mécanique et hydro-mécanique des argiles gonflantes et non gonflantes fortement compactées. *Revue Francaise de Géotechnique* **78** : 31-50.

Bernier, F., Volckaert, G. & Villar, M. 1995. Suction controlled experiments on Boom clay. *International Workshop on Hydro-Thermo-Mechanics of Engineered Clay Barriers and Geological barriers,* Montréal.

Boudali, M., Leroueil, S. and Srinivasa Murthy, B.R. 1994. Viscous behaviour of natural clays. *Proceedings of the 13th International Conference on Soil Mechanics and Foundation Engineering* **1 :** 411-416, New Delhi.

Burghignoli, A., A. Desideri & S. Miliziano 1995. Discussion on Volume change of clays induced by heating as observed in consolidation tests (Towhata and al. 1993). *Soils and Foundations* **35** (3) : 122-124.

Campanella, R.G. & J.K. Mitchell 1968. Influence of temperature variations on soil behavior. *Journal of Soil Mechanics and Foundation Engineering, ASCE* **94** (SM3) : 709-734.

Cui, Y. J. 1993. Etude du comportement d'un limon compacté non saturé et de sa modélisation dans un cadre élasto-plastique. *PhD thesis, Ecole Nationale des Ponts et Chaussées,* 297 p., Paris.

Cui Y. J., Sultan N. & Delage P. 1999. A thermomechanical model for saturated clays. *Canadian Geotechnical Journal*, accepted for publication .

De Bruyn, D. & Thimus, J.F. 1995. The influence of anisotropy on clay strength at high temperature. *Proceedings of the 11$^{th}$ European Conference on Soil Mechanics and Foundation Engineering* **3** : 37-42, Copenhagen.

Decleer, J., Viane, W. and Vandenberghe, N. 1983. Relationships between chemical, physical and mineralogical characteristics of the Rupelian Boom clay, Belgium. *Clay Minerals* **18** : 1-10.

Demars, K.R. and Charles, R.D 1982. Soil volume changes induced by temperature cycling. *Canadian Geotechnical Journal* **19** : 188-194.

Eriksson, L.G. (1989). Temperature effects on consolidation properties of sulphide clays. *Proceedings of the 12th International Conference on Soil Mechanics and Foundation Engineering* **3 :** 2087-2090, Rio de Janeiro.

Gatmiri, B. & Delage P. 1997. A formulation of fully coupled thermal-hydraulic-mechanical behaviour of saturated porous media-Numerical approach. *International Journal for Numerical and Analytical Methods in Geomechanics* **21**: 199-225.

Gray, H. 1936. Progress report on the consolidation of fine-grained soils. *Proceedings of the First International Conference on Soil Mechanics and Foundation Engineering*: 138-141, Cambridge, Mass.

Habibagahi, K. 1977. Temperature effect and the concept of effective void ratio. *Indian Geotechnical Journal,* 14-34.

Hillel D. 1980. Fundamentals of soil physics. *Academic Press,* 413 p.

Houston, S. L. & Lin, H. D. 1987. Thermal consolidation model of pelagic clays. *Marine Geotechnology* **7**: 79-98.

Hueckel T. and Baldi G. 1990. Thermoplasticity of saturated clays: experimental consitutive study. *Journal of Geotechnical Engineering* **116** (12): 1778-1796.

Lambe, T. W. 1960. The structure of compacted clay. *Journal of Soil Mechanics and Foundation Engineering ASCE* **125**: 682-706.

Lewis, R.W., Majorana C.E. & Schrefler B.A. 1986. A coupled finite element model for the consolidation of non-isothermal elastoplastic porous media. *Transport in Porous Media* **1**: 155-178.

Low, P. F. & Lovell, C. W. 1959. The factor of moisture in frost action. *Highway Research Board Bulletin 225*: 23-44.

Mitchell, J.K. 1993. Fundamentals of Soil Behavior. *John Wiley & Sons, Inc.*: 437 p.

Morin, R. & Silva A.J. 1984. The effects of high pressure and high temperature on some physical properties of ocean sediments. *Journal of Geophysical Research* **89**(B1): 511-526.

Neerdael, B., R. Beaufays, M. Buyens, D. De Bruyn & Voet M. 1992. Geomechanical behaviour of Boom clay under ambient and elevated temperature conditions. *Commission of the European Communities, Nuclear Science and Technology EUR 14154*: 108 p.

Noorishad, J., C.F. Tsang & Witherspoon P.A. 1984. Coupled thermal-hydraulic-mechanical phenomena in saturated fractured porous rocks: numerical approach. *Journal of Geophysical Research* **89**: 365-373.

Paaswell, R.E. 1967. Temperature effects on clay soil consolidation. *Journal of Soil Mechanics and Foundation Engineering, ASCE.* **93** (SM3): 9-22.

Picard, J. 1994. Ecrouissage thermique des argiles saturées: application au stockage des déchets radioactifs. Thèse de Doctorat de l'Ecole Nationale des Ponts et Chaussées, 283 p., Paris.

Plum, R.L. and Esrig, M.I. 1969. Some temperature effects on soil compressibility and pore water pressure. Effects of Temperature and Heat on Engineering Behavior of Soils, Special Report 103, *Highway Research Board*: 231-242.

Sultan N. 1997. Etude du comportement thermo-mécanique de l'argile de Boom: expériences et modélisation. *PhD thesis, Ecole Nationale des Ponts et Chaussées,* 310 p., Paris.

Tidfors, M and Sällfors, G. (1989). Temperature effect on preconsolidation pressure. *Geotechnical Testing Journal*; **12**(1), 93-97.

Towhata, I., P. Kuntiwattanakul, I. Seko & Ohishi K. 1993. Volume change of clays induced by heating as observed in consolidation tests. *Soils and Foundations* 33 (4): 170-183.

Towhata, I., P. Kuntiwattanakul, I. Seko & Ohishi K. 1995. Discussion on volume change of clays induced by heating as observed in consolidation tests (Towhata and al. 1993). *Soils and Foundations* 35 (3): 124-127.

Yong, R. T., Taylor L. & Warkentin, B. P. 1962. Swelling pressures of sodium montmorillonite at depressed temperature. *Proceedings of the 11$^{th}$ National Conference on Clays and Clay Minerals*: 268-281.

Youssef, M.S., Sabry, A. and El Ramli A.H. 1961. Temperature changes and their effects on some physical properties of soils. *Proceedings of the Fifth International Conference on Soil Mechanics and Foundation Engineering* **2** : 419-421, Paris.

**TABLE**

Table 1 : Coefficients of consolidation, permeability and compressibility at different temperatures.

|  | 50°C-60°C | 60°C-70°C | 70°C-80°C | 80°C-95°C |
|---|---|---|---|---|
| $c_v$ m²/s | 3.43 10⁻⁸ | 4.32 10⁻⁸ | 4.39 10⁻⁸ | 4.45 10⁻⁸ |
| $k$ m/s | 3.50 10⁻¹² | 3.85 10⁻¹² | 4.2 10⁻¹² | 4.40 10⁻¹² |
| $m_v$ (calculated) | 1.05 10⁻⁵ | 0.91 10⁻⁵ | 0.975 10⁻⁵ | 1.00 10⁻⁵ |
| $m_v$ (measured) | 4.08 10⁻⁵ | 3.70 10⁻⁵ | 3.20 10⁻⁵ | 2.73 10⁻⁵ |

**LIST OF FIGURES**

Figure 1 : Schematic diagram of the triaxial cell.

Figure 2 : Comparison between the two methods of measuring the volume change.

Figure 3 : Thermal volume change and drained water during heating.

Figure 4 : Comparison between the predicted and the measured thermal volumetric strains.

Figure 5 : Effects of the initial void ratio and of the back pressure on the quality of the prediction of drained water

Figure 6 : Thermal consolidation curves, low temperatures (23/30°C, 30/40°C and 40/50°C)

Figure 7 : Thermal consolidation curves, high temperatures (50/60°C, 60/70°C, 70/80°C and 80/95°C).

Figure 8 : Relative volume changes versus logarithm of time, high temperatures.

Figure 9 : Thermal volumetric strains (Boom clay, OCR = 2, p' = 4 MPa)

Figure 10 : Volume of water injected during a permeability test

Figure 11 : Permeability test performed on a sample of Boom clay at various temperatures and stresses.

Figure 12 : A numerical simulation of the thermo-mechanical consolidation of a saturated sand (Noorishad et al. 1984).

Figure 13 : Variation of $c_v$ with temperature.

Figure 14 : Results of the permeability tests, in terms of intrinsic permeability.

Figure 15 : Effect of the plasticity index on the intrinsic permeability of various soils (including the data of Morin and Silva 1984).

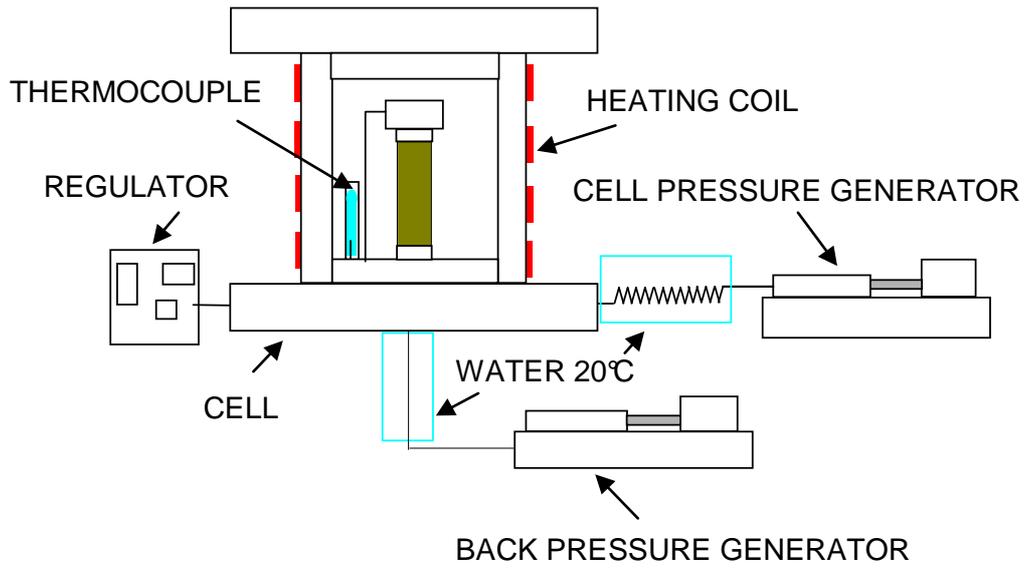

Figure 1: Schematic diagram of the triaxial cell.

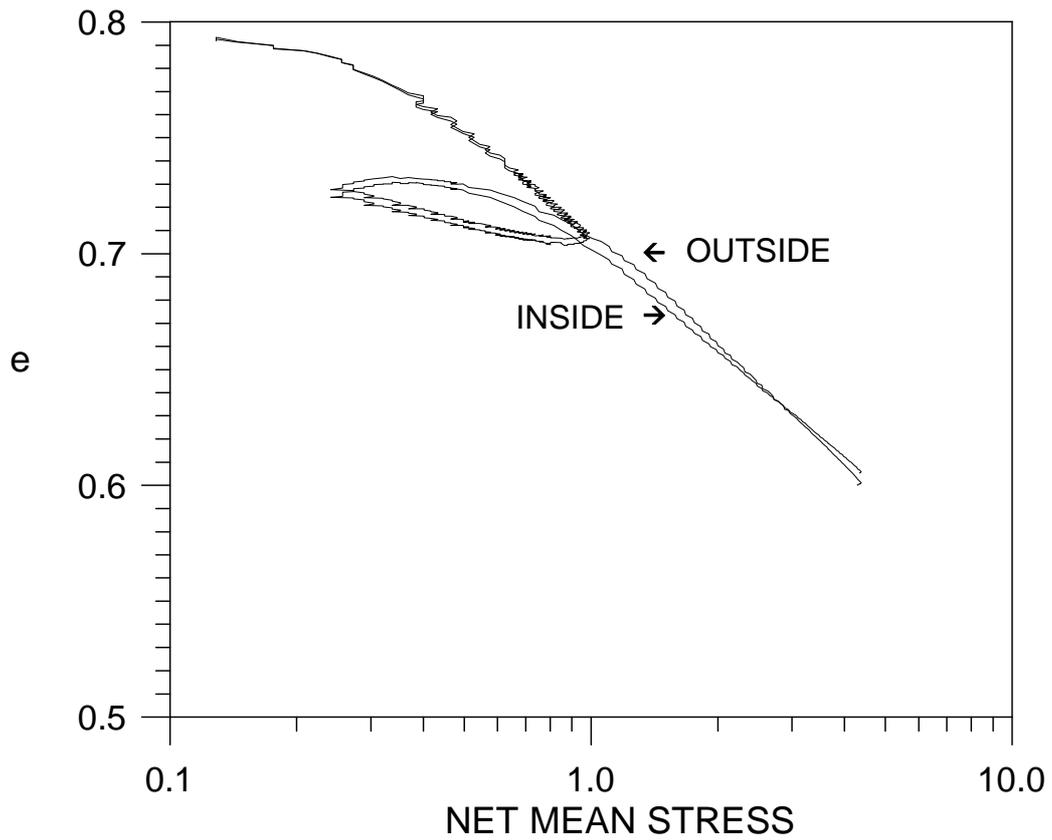

Figure 2: Comparison between the two methods of measuring the volume change.

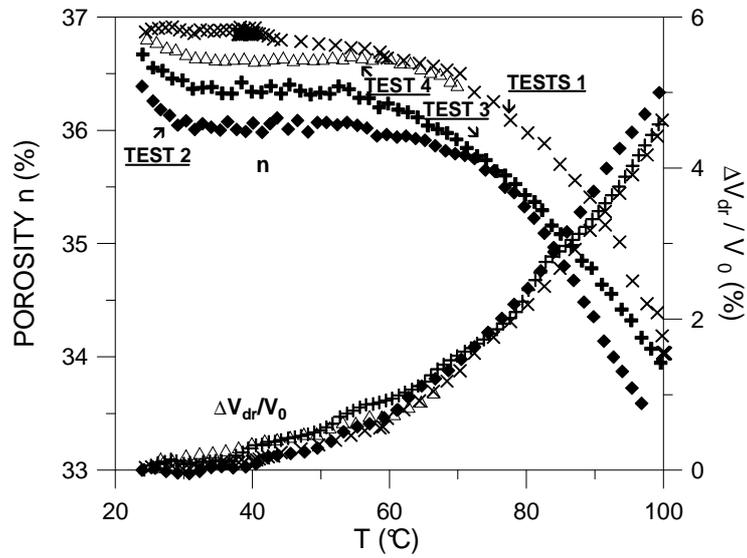

Figure 3 : Thermal volume changes and drained water during heating.

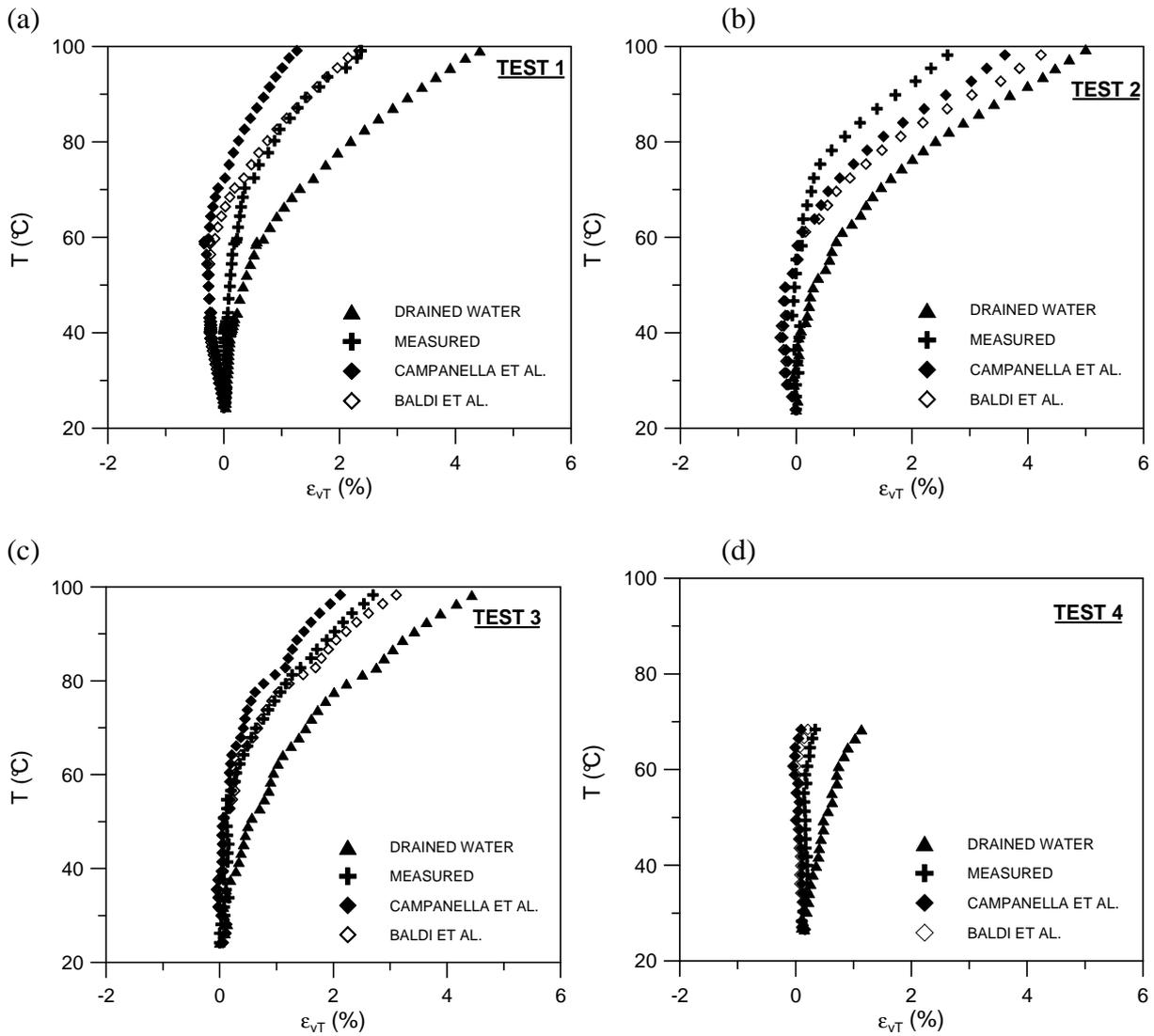

Figure 4 : Comparison between the predicted and the measured thermal volumetric strains.
a) test 1; b) test 2; c) test 3; d) test 4.

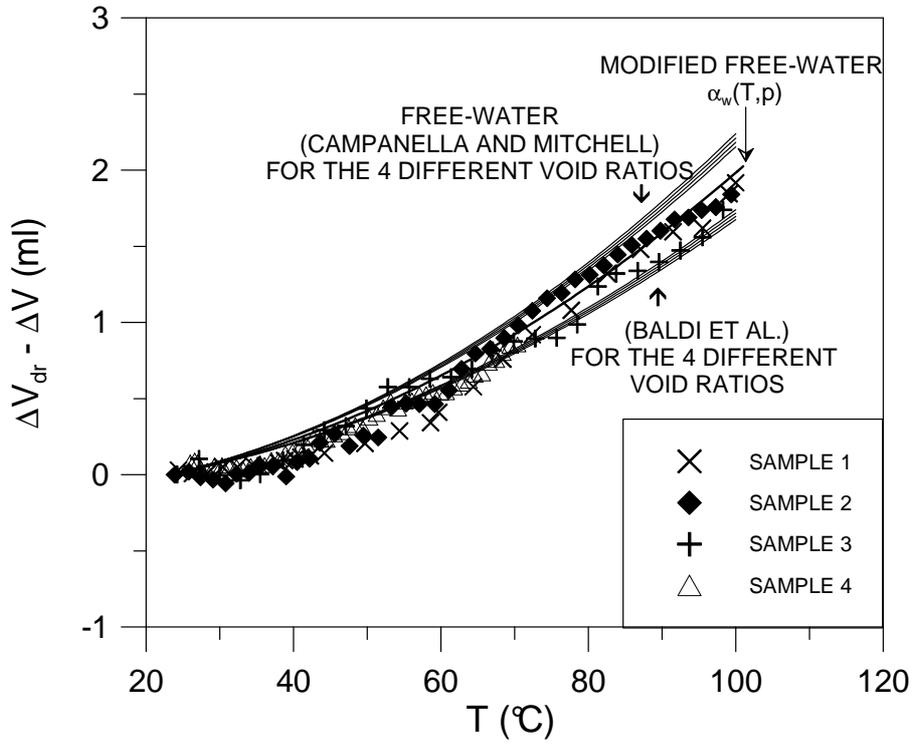

Figure 5 : Effects of initial void ratio and of the back pressure on the quality of the prediction of drained water.

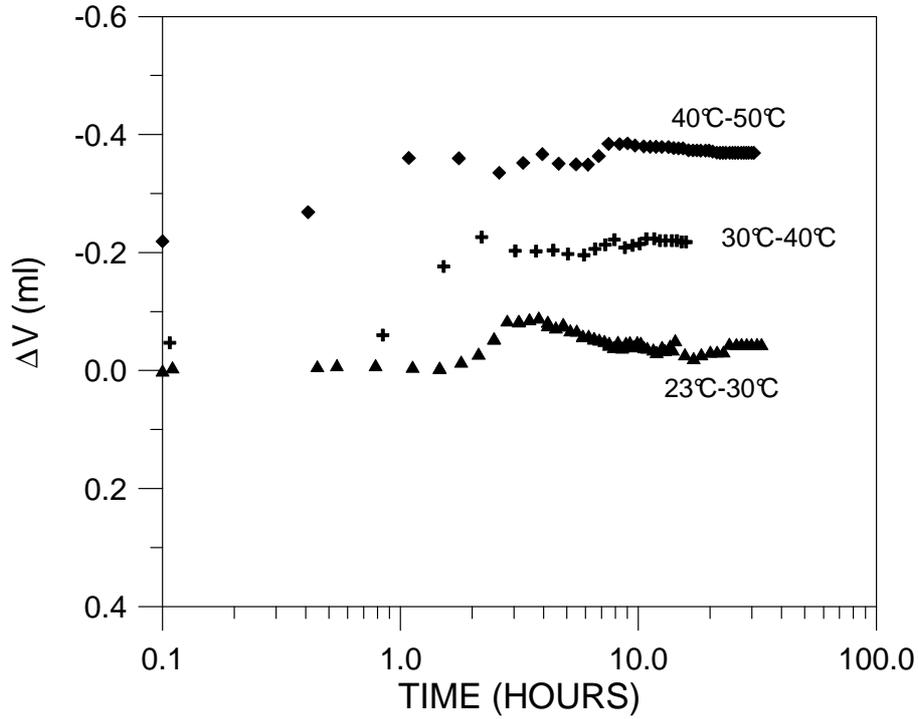

Figure 6 : Thermal consolidation curves, low temperatures (23/30°C, 30/40°C and 40/50°C)

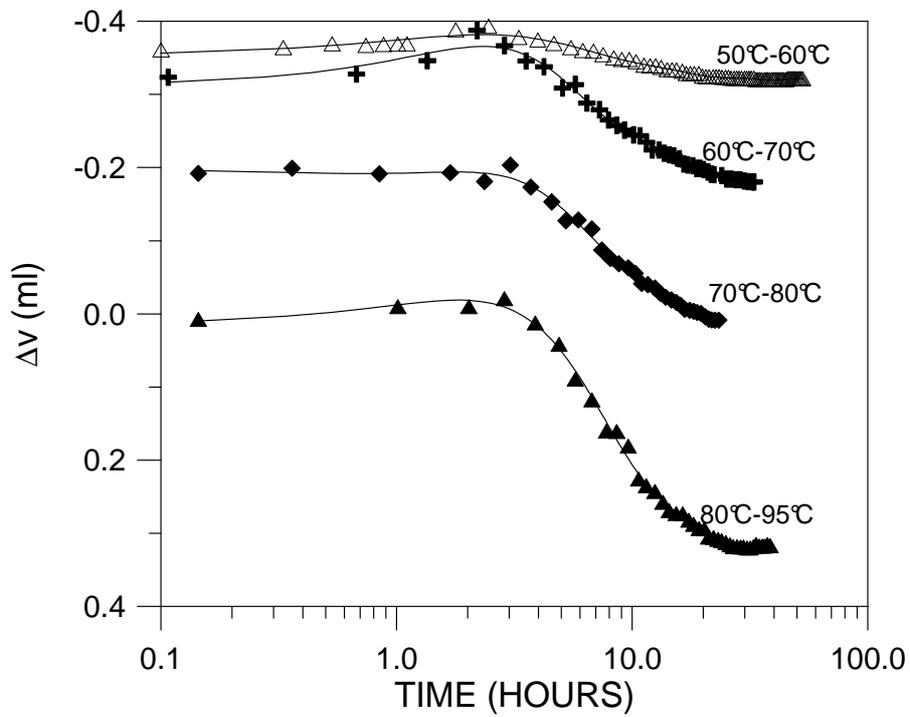

Figure 7 : Thermal consolidation curves, high temperatures (50/60°C, 60/70°C, 70/80°C and 80/95°C).

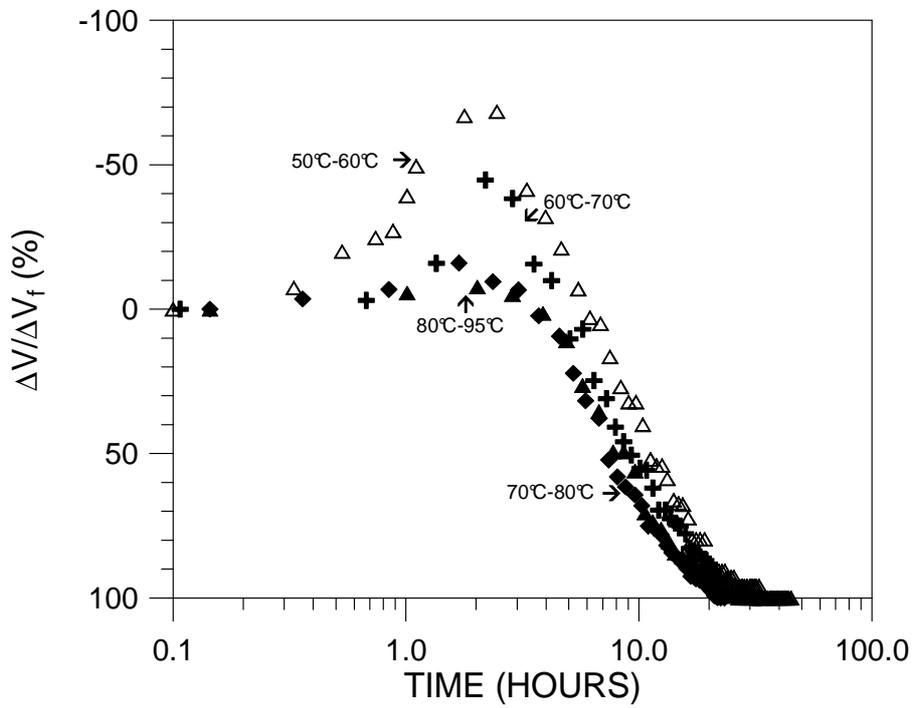

Figure 8 : Relative volume changes versus logarithm of time, high temperatures.

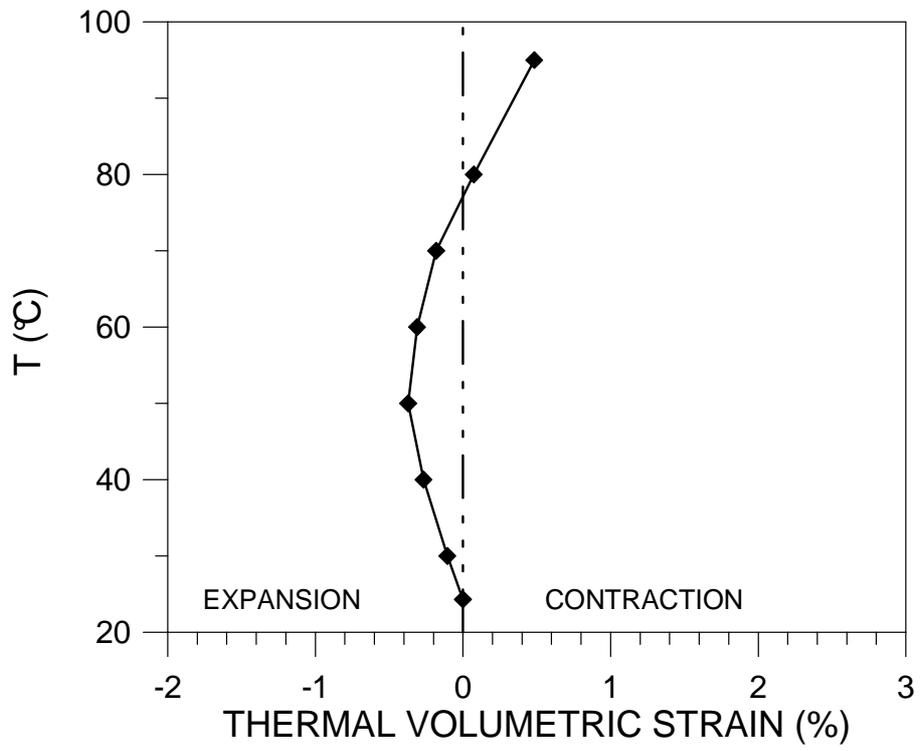

Figure 9 : Thermal volumetric strains (Boom clay, OCR = 2, p' = 4 MPa)

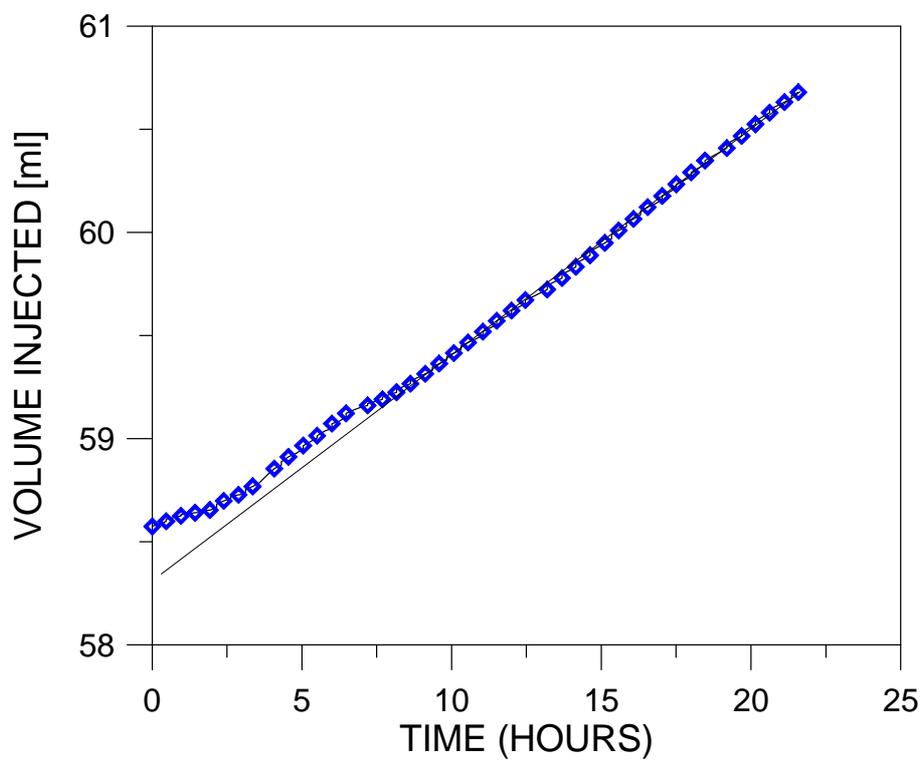

Figure 10 : Volume of water injected during a permeability test

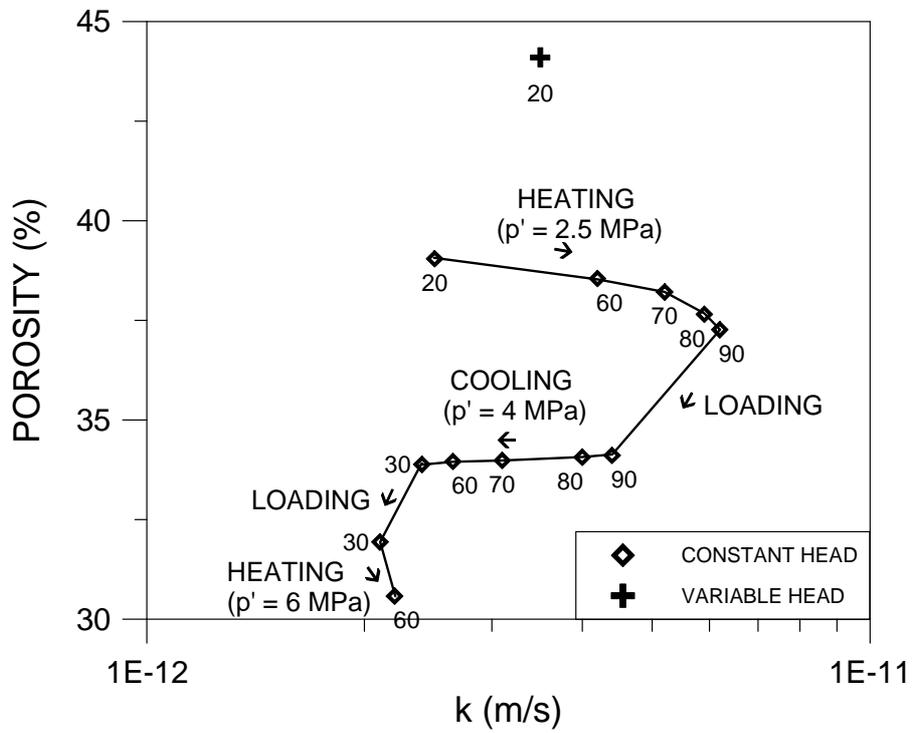

Figure 11 : Permeability tests performed on a sample of Boom clay at various temperatures and stresses.

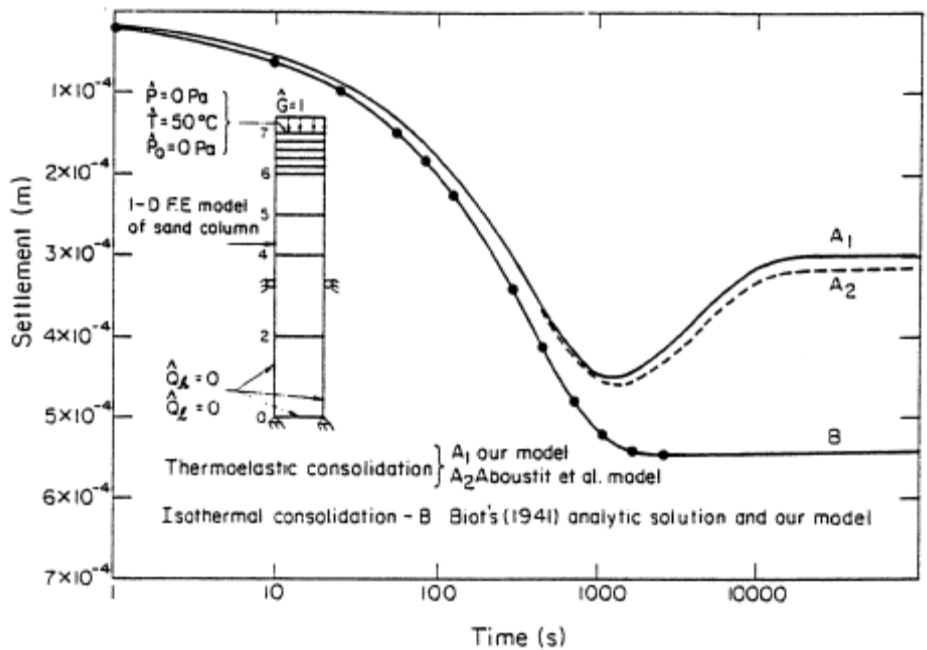

Figure 12 : A numerical simulation of the thermo-mechanical consolidation of a saturated sand (Noorishad et al. 1984).

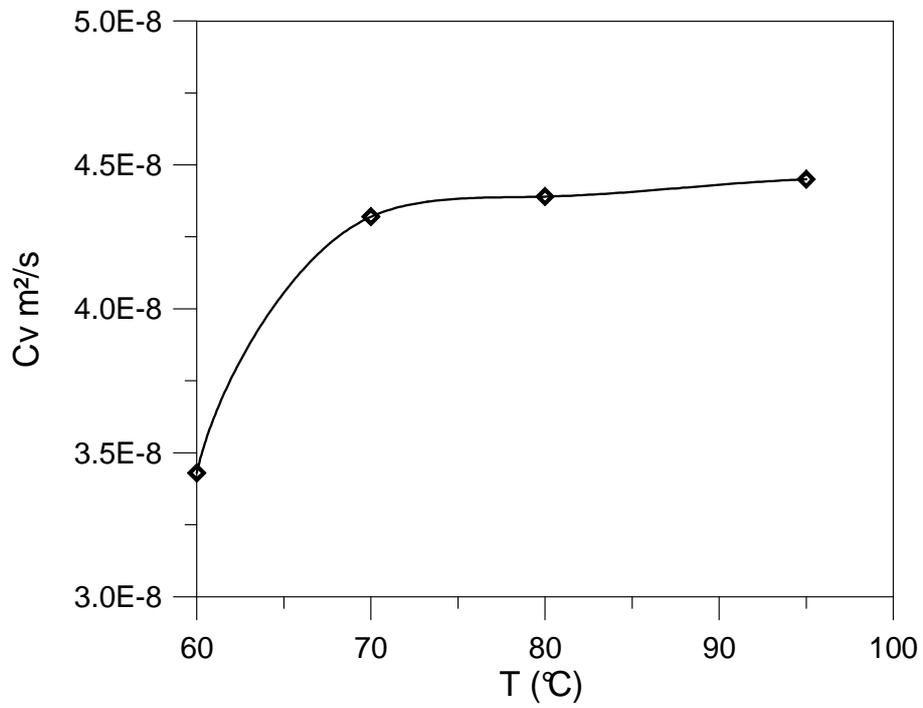

Figure 13 : Variation of cv with temperature.

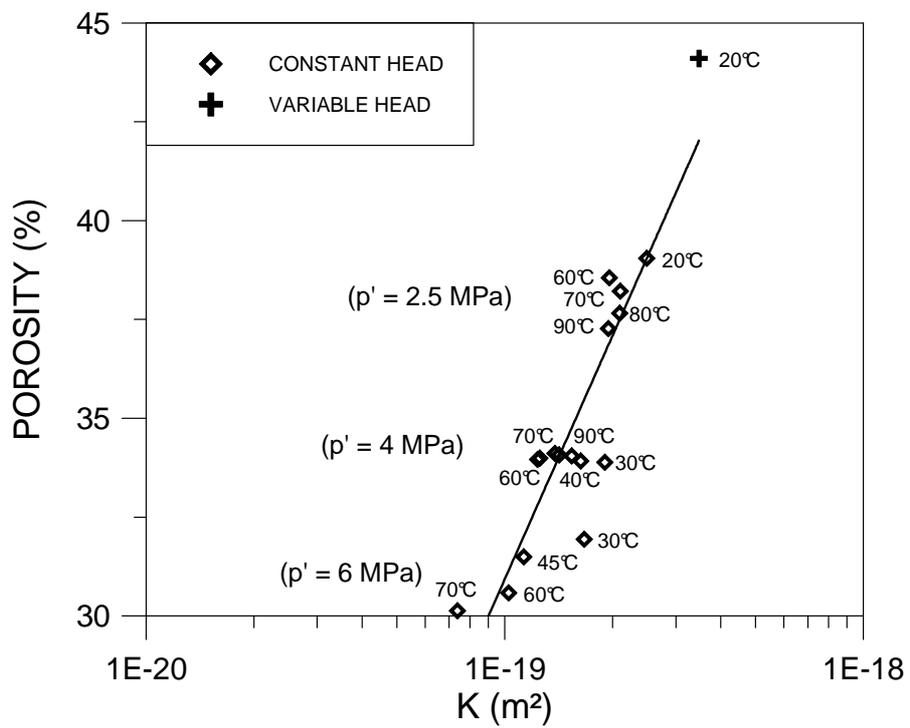

Figure 14 : Results of the permeability tests, in terms of intrinsic permeability.

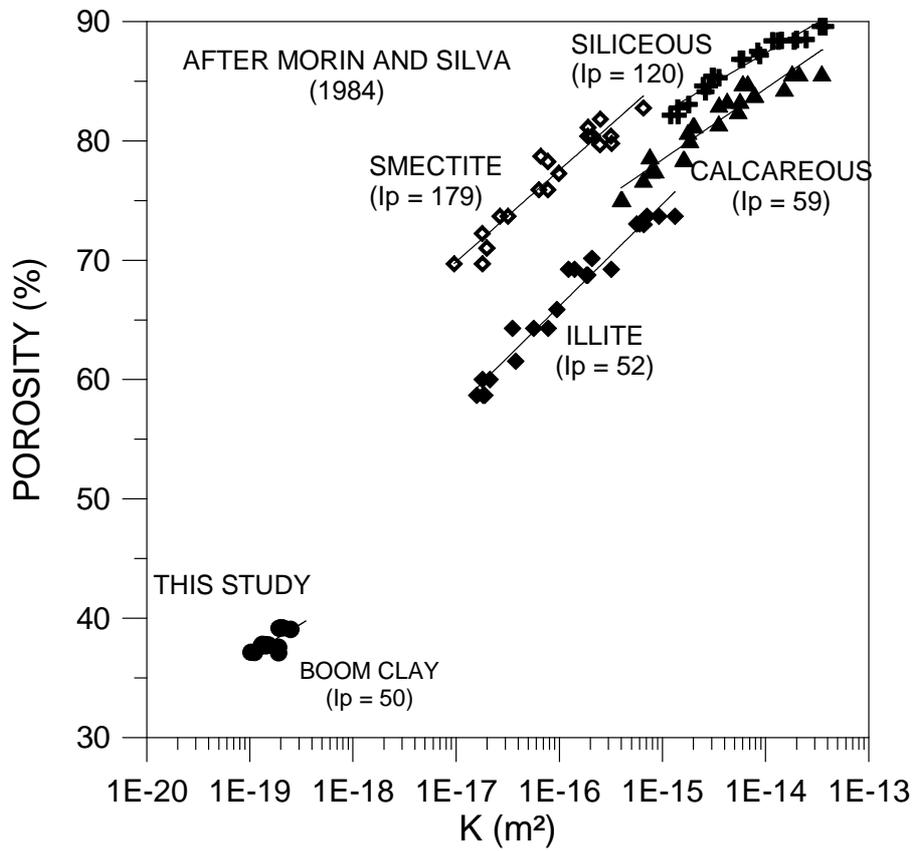

Figure 15: Effect of the plasticity index on the intrinsic permeability of various soils (including the data of Morin and Silva 1984).